\documentclass{article}

\PassOptionsToPackage{numbers,square}{natbib}
\usepackage[final]{neurips_2022}
\usepackage[utf8]{inputenc} % allow utf-8 input
\usepackage[T1]{fontenc}    % use 8-bit T1 fonts
\usepackage{url}            % siimple URL typesetting
\usepackage{booktabs}       % professional-quality tables
\usepackage{amsfonts}       % blackboard math symbols
\usepackage{bm}
\usepackage{nicefrac}       % compact symbols for 1/2, etc.
\usepackage{microtype}      % microtypography
\usepackage{xcolor}         % colors
\definecolor{accent}{RGB}{19,149,186}
\usepackage[colorlinks=true,citecolor=accent,backref=page]{hyperref}       % hyperlinks
\usepackage{graphicx}
\usepackage{tikz}
\usetikzlibrary{
	backgrounds,
    calc,
    positioning,
    scopes,
    shadings,
	shadows,
}

\usepackage{tabularx}
\usepackage{multirow}
\usepackage{makecell}
\usepackage[misc]{ifsym}

\newcommand\mvec[1]{\bm{#1}}
\usepackage{siunitx}
\title{HyperSound: Generating Implicit Neural Representations of Audio Signals with Hypernetworks}

\author{%
    Filip Szatkowski$^{1,2*}$ \quad Karol J. Piczak$^{4*}$ \quad Przemysław Spurek$^{4}$\\
    \quad \textbf{Jacek Tabor}$^{4}$ \quad \textbf{Tomasz Trzciński}$^{1,2,3,4}$\\[0.5em]
    $^1$Warsaw University of Technology \quad $^2$IDEAS NCBR \quad $^3$Tooploox \\
    $^4$Faculty of Mathematics and Computer Science, Jagiellonian University \\[0.5em]  % podobno na UJ jest wymóg podawania wydziału
    \texttt{\{filip.szatkowski.dokt,tomasz.trzcinski\}@pw.edu.pl}\\
    \texttt{\{karol.piczak,przemyslaw.spurek,jacek.tabor\}@uj.edu.pl}\\
}

\begin{document}

\maketitle

\begin{abstract}
    Implicit neural representations (INRs) are a rapidly growing research field, which provides alternative ways to represent multimedia signals. Recent applications of INRs include image super-resolution, compression of high-dimensional signals, or 3D rendering. However, these solutions usually focus on visual data, and adapting them to the audio domain is not trivial. Moreover, it requires a separately trained model for every data sample. To address this limitation, we propose HyperSound, a~meta-learning method leveraging hypernetworks to produce INRs for audio signals unseen at training time. We show that our approach can reconstruct sound waves with quality comparable to other state-of-the-art models.
\end{abstract}

\section{Introduction}\vspace{-.2em}
Implicit neural representations (INRs) are coordinate-based representations of multimedia signals, where the signal is modeled with a neural network. Such representations are decoupled from the spatial resolution, so the signal can be resampled at any arbitrary frequency. At the same time, the memory requirements for its storage remain constant. The field of INRs is rapidly growing, and their applications include super-resolution~\cite{sitzmann2020implicit,mehta2021modulated}, compression~\cite{sitzmann2020implicit,mehta2021modulated} or 3D rendering~\cite{mildenhall2021nerf}. However, in the audio domain, the evaluation of these approaches was so far limited to learning individual INRs of particular input recordings~\cite{sitzmann2020implicit}, which is highly ineffective.

Meta-learning methods, such as hypernetworks~\cite{ha2016hypernetworks}, can generate INRs for any arbitrary signal with a single model. Hypernetworks learn to generate weights for smaller target networks, which can serve as INRs. Hypernetworks were successfully applied in obtaining INRs for images~\cite{mehta2021modulated,klocek2019hypernetwork}, point clouds~\cite{mehta2021modulated,spurek2020hypernetwork} and videos~\cite{mehta2021modulated}. 
However, creating INRs for high-dimensional and high-variance data such as audio is difficult, and training hypernetworks for this task can be unstable.
%However, creating INRs for high-dimensional and high-variance signals such as audio is difficult. The same reasons make training hypernetworks for this task challenging, as it requires a lot of computational resources and can be unstable.

In this work, we propose a meta-learning approach based on hypernetworks, where we learn a~general recipe for creating INRs of arbitrary audio samples from outside of the training dataset. To our knowledge, our model is the first application of hypernetwork-based INRs to the audio domain.

\begin{figure}[!h]
    \centering
    \includegraphics[width=\textwidth]{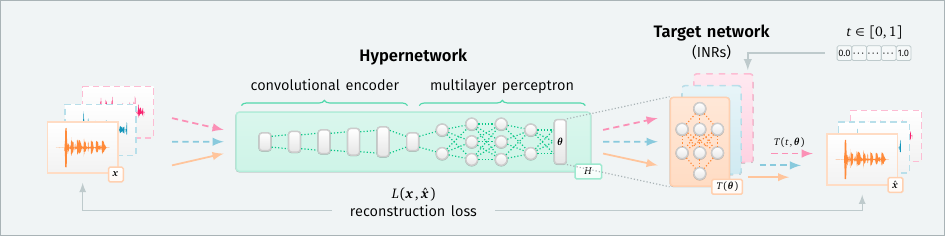}
    \caption{Overview of the HyperSound framework. We use a~single hypernetwork model to produce distinct INRs based on arbitrary audio signals provided as input.\vspace*{-.7em}} 
    \label{fig:teaser}
\end{figure}

\section{Related Works}\vspace{-.2em}

% \textbf{Implicit neural representations (INRs)} are applied across many multimedia domains.
% \textbf{NeRF}~\cite{mildenhall2021nerf} based INRs are currently state-of-the-art for image-based rendering and view synthesis. \textbf{SIREN}~\cite{sitzmann2020implicit} shows how to produce good quality INRs for various signals, such as image, video and sound signals, using neural networks with periodic activation functions and dedicated weight initialization scheme. \textbf{Modulated Periodic Activations}~\cite{mehta2021modulated} further improve the quality of SIREN-based INRs for high resolution signals by introducing modulation and synthesis sub-networks. However, in the audio domain, the evaluation of these approaches was so far limited to learning individual INRs of particular input recordings.

\textbf{Implicit neural representations (INRs)} are applied across many multimedia domains.
\textbf{NeRF}~\cite{mildenhall2021nerf}-based INRs are currently state-of-the-art for image-based rendering and view synthesis. \textbf{SIREN}~\cite{sitzmann2020implicit} shows how to produce good quality INRs for various signals, such as image, video, and sound signals, using neural networks with periodic activation functions and dedicated weight initialization scheme. \textbf{Modulated Periodic Activations}~\cite{mehta2021modulated} further improve the quality of SIREN-based INRs for high-resolution signals by introducing modulation and synthesis sub-networks. However, in the audio domain, the evaluation of these approaches was so far limited to learning individual INRs of particular input recordings.

\textbf{Hypernetworks}~\cite{ha2016hypernetworks} are a meta-learning framework, where one network (hypernetwork) generates the weights for another network (target network). Hypernetworks can be used for a variety of tasks such as model compression~\cite{zhao2020meta}, continual learning~\cite{von2019continual} or generating \textbf{INRs}. In particular, hypernetworks were used to generate INRs for images~\cite{klocek2019hypernetwork}, shapes~\cite{spurek2020hypernetwork} and videos~\cite{mehta2021modulated}. To our knowledge, our work is the first application of hypernetworks for audio INRs generation.

First successful attempts at \textbf{raw waveform processing with deep neural networks} were models such as \textbf{WaveNet}~\cite{oord2016wavenet} and \textbf{SampleRNN}~\cite{mehri2016samplernn}, but their autoregressive nature makes them slow and prone to accumulation of errors. Later architectures such as \textbf{ParallelWaveNet}~\cite{oord2018parallel}, \textbf{NSynth}~\cite{engel2017neural}, \textbf{MelGAN}~\cite{kumar2019melgan} or \textbf{SING}~\cite{defossez2018sing} proposed non-autoregressive architectures for audio generation. Recent autoencoder-based models such as \textbf{RAVE}~\cite{caillon2021rave} or \textbf{SoundStream}~\cite{zeghidour2021soundstream} are able to process high-resolution signals in an end-to-end fashion, producing audio of very good perceptual quality.

\section{Model overview}\vspace{-.2em}
\label{sec:3}
Sound waves are traditionally represented digitally as a collection of amplitude values sampled at regular intervals, which approximates a~continuous real function $x(t)$. Our goal is to obtain a~meta-recipe for generating audio INRs replicating such functions. While creating INRs for particular audio samples can be quite easily done with gradient descent, finding a general solution is much harder due to the inherent complexity of audio time series. Therefore, similar to \cite{klocek2019hypernetwork}, we model such functions with neural networks $T$ (target networks), parameterized by weights generated by another neural network $H$ (hypernetwork). Our framework, shown in Fig.~\ref{fig:teaser}, can be described as
\begin{equation}
    \mvec{\theta}_{\mvec{x}} = H(\mvec{x}),\vspace*{-.3em}
\end{equation}\begin{equation}
    \hat{x}(t) = T(t, {\mvec{\theta}}_{\mvec{x}}).
\end{equation}

\vspace*{-.8em}
\subsection{Hypernetwork architecture}\vspace{-.2em}
Typical audio recordings contain several thousands of samples, so the hypernetwork is composed of a convolutional encoder that produces a~latent representation of a~lower dimensionality, and fully connected layers that transform this representation to weights $\mvec{\theta}$ of the target network.

We use an encoder based on SoundStream~\cite{zeghidour2021soundstream}, and the fully connected part of the hypernetwork is composed of six fully-connected layers with biases and ELU~\cite{clevert2015fast} activation, where the last layer produces the flattened weights of the target network.

\subsection{Approximating sound waves with neural networks}\vspace{-.2em}
\label{target_network}
A target network should be as small as possible to avoid overfitting and explosion of weights in the hypernetwork, but must be expressive enough to represent a wide variety of audio recordings. Our target network has one input and one output, and consists of a~positional embedding layer followed by four fully-connected layers of \num{256} neurons with biases and ReLU activation. Inspired by NeRF~\cite{mildenhall2021nerf}, we define embedding vectors $\mvec{\gamma}$ as
\begin{equation}
    \mvec{\gamma}(t) = \left[\sin(2^0 \pi t),\, \cos(2^0 \pi t),\, \sin(2^1 \pi t),\, \cos(2^1 \pi t),\, \ldots,\, \sin({2}^{L-1} \pi t),\, \cos({2}^{L-1} \pi t)\right],
\end{equation}
where $t$ is the time coordinate and $L$ denotes the embedding size. We rescale input coordinates to a~range of $[0, 1]$ and use $L=16$.

\subsection{Optimization}\vspace{-.2em}
\label{sec:optim}
We train the hypernetwork in a supervised fashion using backpropagation. To obtain audio results that are more perceptually pleasant, we use a~loss function that penalizes the reconstruction error both in the time and frequency domains. Given an original recording $\mvec{x}$ and its reconstruction $\hat{\mvec{x}}$ generated with a~target network, we compute the loss function as
\begin{equation}
    \label{eq_loss}
    L(\mvec{x}, \hat{\mvec{x}}) = {\lambda}_{SL1} * L_{SL1}(\mvec{x}, \hat{\mvec{x}}) + {\lambda}_{STFT} * L_{STFT}(\mvec{x}, \hat{\mvec{x}}),
\end{equation}
where $ L_{SL1}$ is a~smooth L1 loss~\cite{girshick2015fast} with $\beta=0.1$, $L_{STFT}$ is a~multi-resolution mel-scale STFT loss introduced in ParallelWaveGAN~\cite{yamamoto2020parallel} and ${\lambda}_{SL1}$, ${\lambda}_{STFT}$ are the weights of these two losses. We use ${\lambda}_{SL1}=1$ and ${\lambda}_{STFT}=1$. For STFT loss, we use \num{128} mel bins and FFT sizes of $[512, 1024, 2048]$ with matching window sizes and an overlap of \num{87.5}\%.

\section{Experiments}\vspace{-.2em}

\begin{figure}[!t]
    \begin{center}
    \begin{tikzpicture}[%
        scale=1.0,
        node distance=0.5cm,
        on grid,
        % background rectangle/.style={draw=black,fill=red},
        % show background rectangle,
        inner frame sep=-0.01cm,
    ]
        \clip (0, 0) rectangle (\textwidth, 3.8);
        \sffamily
        \node [inner sep=0cm, anchor=south west] at (2.5, 0) {\includegraphics[height=3.8cm,width=11cm,trim={0 0 0 0},clip]{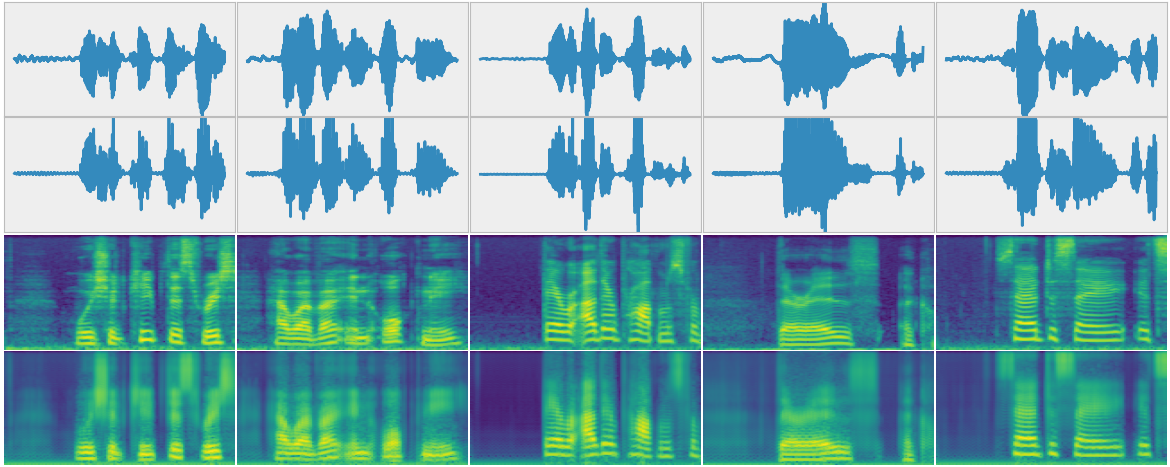}};
        \node [align=left, anchor=west] at (0.25, 3.325) {\tiny \textbf{Ground truth}};
        \node [align=left, anchor=west] at (0.25, 2.375) {\tiny \textbf{Reconstruction}};
        \node [align=left, anchor=west] at (0.25, 1.425) {\tiny \textbf{Ground truth}};
        \node [align=left, anchor=west] at (0.25, 0.475) {\tiny \textbf{Reconstruction}};
    \end{tikzpicture}
    \end{center}
    \vspace*{-.5em}
    \caption{Examples of VCTK validation samples reconstructed with HyperSound.}
    \vspace{-.8em}
    \label{reconstructions}
\end{figure}
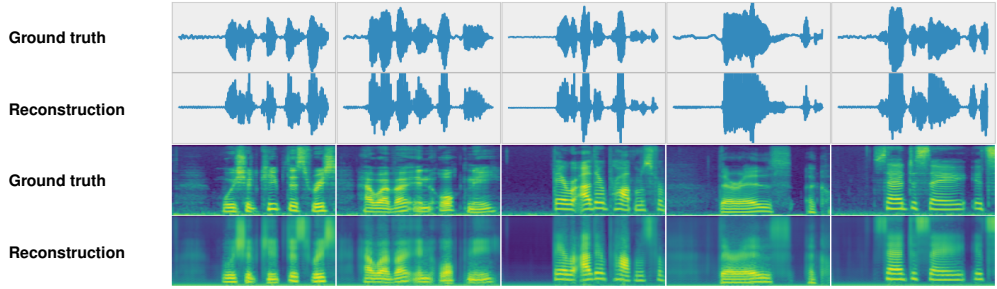

We test the reconstruction quality of our model on the VCTK dataset downsampled to $f=\SI{22050}{Hz}$, with the recordings of the last \num{10} speakers retained as a validation set.
%This split results in a training dataset of 32796 examples and a validation set composed of 2022 examples. 
We set the recording length to \num{32768} samples, and use data augmentations such as random crop, phase mangle, or dequantization proposed in RAVE~\cite{caillon2021rave}. We train the models for 1.25M steps using the AdamW optimizer~\cite{loshchilov2017decoupled} with a~learning rate of \num{5e-5} and a~batch size of \num{16}. In Fig.~\ref{reconstructions}, we show waveforms and spectrograms of the reconstructions obtained with our model on samples selected from the validation set.

Since there is no consensus on a~single approach for quantitative evaluation of audio quality, we assess the reconstruction results with multiple metrics such as MSE, Log-Spectral Distance (LSD)~\cite{liu2022neural}, SI-SNR~\cite{luo2018tasnet}, PESQ~\cite{rix2001perceptual}, STOI~\cite{taal2010short} and CDPAM~\cite{manocha2021cdpam}. We also compare the reconstruction quality of our model with the RAVE baseline. Moreover, we test the quality of resampling performed by our model. Finally, we investigate the impact of the target network size and the employed loss function on the reconstruction quality. The results of our experiments are shown in Tab.~\ref{table:results}. Unless explicitly mentioned, model hyperparameters in experiments are as described in Sec.~\ref{sec:3}.

\textbf{Comparision with RAVE.} We compare the reconstruction error of HyperSound with RAVE~\cite{caillon2021rave} trained for 3M steps on the same dataset, with an identical sampling rate of~\SI{22050}{Hz}. We find that our model produces reconstructions closer to the originals in the spectral domain and obtains better perceptual scores of PESQ and STOI. However, RAVE reconstructions achieve better MSE, SI-SNR, and perceptual CDPAM scores, with slightly less perceptible noise and robotic artifacts.

\textbf{Reconstruction quality while resampling.} We also test the reconstruction error of the model trained with a sampling rate of \SI{22050}{Hz} when used for downsampling and upsampling \SI{22050}{Hz} recordings to other commonly used frequencies: \SI{8000}{Hz}, \SI{16000}{Hz} and \SI{44100}{Hz}. We obtain the ground truth by downsampling the original VCTK \SI{48}{kHz} recordings to the desired sampling frequency using \textit{soxr} high-quality setting. As shown by the LSD levels, the hypernetwork approach works best when reconstructing signals close to the sampling rate used for training. Nevertheless, it does not collapse when queried with time coordinates from outside the training domain. However, we hypothesize that for proper super-resolution capabilities, our learning regime would require an introduction of an additional loss term directed specifically at the quality of representation in higher frequency bands. 

\textbf{Impact of target network architecture.} We compare our model which uses the baseline target network (4 layers of \num{256} neurons each, 206K parameters) with target networks that use \num{4} layers of \num{64} neurons (14K parameters) and \num{6} layers of \num{384} neurons (752K parameters). We find that the base version presents an optimal trade-off between the reconstruction quality and computational requirements, as the size of the last hypernetwork layer scales linearly with the number of parameters in the target network. The smaller variant of the target network achieves comparable results to the base using fewer parameters than the number of samples in the original signal.

\textbf{Impact of the loss function.} We compare HyperSound trained with the loss function as described in Sec.~\ref{sec:optim} with models trained using the STFT loss where we do not apply the mel-scale, but use a wider array of FFT sizes of $[128, 256, 512, 1024, 2048]$. We also try training without Smooth L1 loss (${\lambda}_{SL1}=0, {\lambda}_{STFT}=1$). Moreover, we find that STFT is vital for stable training and obtaining perceptually plausible reconstructions, as our training runs with ${\lambda}_{SL1}=1.$ and ${\lambda}_{STFT}=[0.1,0.01]$ collapsed. As evidenced by the results, the L1 part of the loss function leads to slightly better results and faster training, as it enables the model to correctly learn the DC offset. 

\begin{table}[!t]
\setlength\lightrulewidth{0.1pt}
\footnotesize
\caption{HyperSound evaluation based on reconstructions of the VCTK validation set}\vspace*{-.7em}
\label{table:results}
\vspace*{1.0em}
\begin{tabularx}{\textwidth}{Xlrrrrrr}
    %\multicolumn{8}{c}{\textbf{Impact of target network model size}}\\[0.3em]
    % \toprule\\[-0.95em]
        & & \textbf{MSE} & \textbf{LSD} & \textbf{SI-SNR} & \textbf{PESQ} & \textbf{STOI} & \textbf{CDPAM} \\
        & \textit{Ideal metric behavior} & $\to 0$ & $\to 0$ & $\to 100$ & $\to 4.5$ & $\to 1$ & $\to 0$ \\
    \midrule\\[-0.9em]
    \multirow{3}{*}{\makecell[l]{\textbf{Comparison}\\\textbf{with RAVE}}}  & \textbf{Model} & \multicolumn{6}{c}{} \\[0.25em]
    %\midrule\\[-0.9em]
        % \cmidrule{2-8}
        & \textit{HyperSound}$^*$ & 0.049 & \textbf{0.99} & -23.50 & \textbf{1.75} &\textbf{ 0.87} & 0.35 \\
        & RAVE & \textbf{0.040} & 1.19 & \textbf{-22.42} & 1.22 & 0.74 & \textbf{0.19} \\
    \midrule\\[-0.9em]
    %\bottomrule\\[1.0em]
    %\multicolumn{8}{c}{\textbf{Results for resampling with hypernetwork}}\\[0.3em]
    %\toprule\\[-0.95em]
    \multirow{5}{*}{\makecell[l]{\textbf{Reconstruction}\\\textbf{quality while}\\\textbf{resampling}}} & \textbf{Target SR~[Hz]} & \multicolumn{6}{c}{} \\[0.25em]
    %\midrule\\[-0.9em]
        % \cmidrule{2-8}
        & \num{8000} & 0.048 & 1.24 & -23.76 & 1.67 & 0.83 & 0.37 \\
        & \num{16000} & \textbf{0.047} & 1.19 & -23.80 & 1.66 & 0.86 & \textbf{0.35 }\\
        & \textit{22 050}$^*$ & 0.049 & \textbf{0.99} & \textbf{-23.50} & \textbf{1.75} & \textbf{0.87} &\textbf{ 0.35}\\
        & \num{44100} & \textbf{0.047} & 1.38 & -24.16 & 1.72 & \textbf{0.87} & 0.38 \\
    \midrule\\[-0.9em]
    %\bottomrule\\[1.0em]
    \multirow{4}{*}{\makecell[l]{\textbf{Impact of target}\\\textbf{network size}}} & \textbf{Network size} & \multicolumn{6}{c}{} \\[0.25em]
    % \cmidrule{2-8}
    %\midrule\\[-0.9em]
        & Small [4 x 64] & 0.048 & 1.07 & -24.56 & 1.56 & 0.84 & \textbf{0.35} \\  % 14 657 params
        & \textit{Base [4 x 256]}$^*$ & 0.049 & 0.99 & \textbf{-23.50} & 1.75 & 0.87 & \textbf{0.35} \\  % 206 081 params
        & Large [6 x 384] & \textbf{0.047} & \textbf{0.96} & -23.70 & \textbf{1.86} & \textbf{0.89} & 0.36 \\  % 752 257 params
    \midrule\\[-0.9em]
    %\bottomrule\\[1.0em]
    %\multicolumn{8}{c}{\textbf{Impact of the loss function}}\\[0.3em]
    %\toprule\\[-0.95em]
    \multirow{5}{*}{\makecell[l]{\textbf{Impact of the}\\\textbf{loss function}}} & \textbf{Loss function} & \multicolumn{6}{c}{} \\[0.25em]
    %\midrule\\[-0.9em]
        % \cmidrule{2-8}
        & L1$\,+\,$STFT & 0.052 & \textbf{0.94} & -25.05 & 1.35 & 0.81 & 0.25 \\ %https://wandb.ai/hypersound/hypersound/runs/2j285dgq/overview?workspace=user-fszatkowski
        & Only STFT & 0.055 & \textbf{0.94} & -25.63 & 1.34 & 0.81 & \textbf{0.23} \\ % https://wandb.ai/hypersound/hypersound/runs/2y1jj0uk/overview?workspace=user-fszatkowski
        & \textit{L1$\,+\,$MelSTFT}$^*$ & \textbf{0.049} & 0.99 & -23.50 &\textbf{ 1.75} & \textbf{0.87} & 0.35 \\ % https://wandb.ai/hypersound/hypersound/runs/3lc1h8as
        & Only MelSTFT & 0.050 & 1.02 & \textbf{-23.31} & 1.70 &\textbf{ 0.87 }& 0.29 \\ % https://wandb.ai/hypersound/hypersound/runs/1airbhw3/overview?workspace=user-fszatkowski
        % & L1+MelWinSTFT & 0. & 0. & 0. & 0. & 0. & 0.  \\ % https://wandb.ai/hypersound/hypersound/runs/2fucdd6s?workspace=user-fszatkowski
    % \midrule\\[-0.9em]
    \midrule\\[-0.9em]
    %\multicolumn{8}{c}{\textbf{Comparision with RAVE}}\\[0.3em]
    %\toprule\\[-0.95em]
    % TODO: Table formatting, verify SI-SNR
    % \bottomrule\\[1.0em]
    \multicolumn{8}{r}{$^*$\textit{default HyperSound hyperparameters}}\\
    % \bottomrule\\[1.0em]
\end{tabularx}
\vspace{-1em}
\end{table}%

\section{Conclusion}
We demonstrate the possibility of applying hypernetworks to the generation of implicit neural representations for audio signals. Reconstructions generated with our model are quantitatively comparable to the state-of-the-art model RAVE. However, we find that the perceptual quality of our reconstructions is still slightly lacking. We hope that our work can be further improved by optimizing the hypernetwork architecture and designing target networks better suited to the audio domain. Initial results for signal compression are also promising but require further investigation.

\begin{ack}
This work was supported by Foundation for Polish Science (grant no POIR.04.04.00-00-14DE/18-00) carried out within the Team-Net program co-financed by the European Union under the European Regional Development Fund, as well as the National Centre of Science (Poland) Grant No. 2020/39/B/ST6/01511. Przemysław Spurek is supported by the National Centre of Science (Poland) Grant No. 2021/43/B/ST6/01456.
\end{ack}

{
\small
\bibliography{bibfile}

\begin{thebibliography}{25}
\providecommand{\natexlab}[1]{#1}
\providecommand{\url}[1]{\texttt{#1}}
\expandafter\ifx\csname urlstyle\endcsname\relax
  \providecommand{\doi}[1]{doi: #1}\else
  \providecommand{\doi}{doi: \begingroup \urlstyle{rm}\Url}\fi

\bibitem[Sitzmann et~al.(2020)Sitzmann, Martel, Bergman, Lindell, and
  Wetzstein]{sitzmann2020implicit}
Vincent Sitzmann, Julien Martel, Alexander Bergman, David Lindell, and Gordon
  Wetzstein.
\newblock {Implicit} {Neural} {Representations} with {Periodic} {Activation}
  {Functions}.
\newblock \emph{Advances in Neural Information Processing Systems},
  33:\penalty0 7462--7473, 2020.

\bibitem[Mehta et~al.(2021)Mehta, Gharbi, Barnes, Shechtman, Ramamoorthi, and
  Chandraker]{mehta2021modulated}
Ishit Mehta, Micha{\"e}l Gharbi, Connelly Barnes, Eli Shechtman, Ravi
  Ramamoorthi, and Manmohan Chandraker.
\newblock Modulated {Periodic} {Activations} for {Generalizable} {Local}
  {Functional} {Representations}.
\newblock In \emph{Proceedings of the IEEE/CVF International Conference on
  Computer Vision}, pages 14214--14223, 2021.

\bibitem[Mildenhall et~al.(2021)Mildenhall, Srinivasan, Tancik, Barron,
  Ramamoorthi, and Ng]{mildenhall2021nerf}
Ben Mildenhall, Pratul~P Srinivasan, Matthew Tancik, Jonathan~T Barron, Ravi
  Ramamoorthi, and Ren Ng.
\newblock {NeRF}: {Representing} {Scenes} as {Neural} {Radiance} {Fields} for
  {View} {Synthesis}.
\newblock \emph{Communications of the ACM}, 65\penalty0 (1):\penalty0 99--106,
  2021.

\bibitem[Ha et~al.(2016)Ha, Dai, and Le]{ha2016hypernetworks}
David Ha, Andrew Dai, and Quoc~V Le.
\newblock {HyperNetworks}.
\newblock \emph{arXiv preprint arXiv:1609.09106}, 2016.

\bibitem[Klocek et~al.(2019)Klocek, Maziarka, Wo{\l}czyk, Tabor, Nowak, and
  {\'S}mieja]{klocek2019hypernetwork}
Sylwester Klocek, {\L}ukasz Maziarka, Maciej Wo{\l}czyk, Jacek Tabor, Jakub
  Nowak, and Marek {\'S}mieja.
\newblock Hypernetwork functional image representation.
\newblock In \emph{International Conference on Artificial Neural Networks},
  pages 496--510. Springer, 2019.

\bibitem[Spurek et~al.(2020)Spurek, Winczowski, Tabor, Zamorski, Zi{\k{e}}ba,
  and Trzci{\'n}ski]{spurek2020hypernetwork}
Przemys{\l}aw Spurek, Sebastian Winczowski, Jacek Tabor, Maciej Zamorski,
  Maciej Zi{\k{e}}ba, and Tomasz Trzci{\'n}ski.
\newblock Hypernetwork approach to generating point clouds.
\newblock \emph{arXiv preprint arXiv:2003.00802}, 2020.

\bibitem[Zhao et~al.(2020)Zhao, von Oswald, Kobayashi, Sacramento, and
  Grewe]{zhao2020meta}
Dominic Zhao, Johannes von Oswald, Seijin Kobayashi, Jo{\~a}o Sacramento, and
  Benjamin~F Grewe.
\newblock Meta-{Learning} via {Hypernetworks}.
\newblock 2020.

\bibitem[Von~Oswald et~al.(2019)Von~Oswald, Henning, Sacramento, and
  Grewe]{von2019continual}
Johannes Von~Oswald, Christian Henning, Jo{\~a}o Sacramento, and Benjamin~F
  Grewe.
\newblock Continual learning with hypernetworks.
\newblock \emph{arXiv preprint arXiv:1906.00695}, 2019.

\bibitem[Oord et~al.(2016)Oord, Dieleman, Zen, Simonyan, Vinyals, Graves,
  Kalchbrenner, Senior, and Kavukcuoglu]{oord2016wavenet}
Aaron van~den Oord, Sander Dieleman, Heiga Zen, Karen Simonyan, Oriol Vinyals,
  Alex Graves, Nal Kalchbrenner, Andrew Senior, and Koray Kavukcuoglu.
\newblock {WaveNet}: {A} {Generative} {Model} for {Raw} {Audio}.
\newblock \emph{arXiv preprint arXiv:1609.03499}, 2016.

\bibitem[Mehri et~al.(2016)Mehri, Kumar, Gulrajani, Kumar, Jain, Sotelo,
  Courville, and Bengio]{mehri2016samplernn}
Soroush Mehri, Kundan Kumar, Ishaan Gulrajani, Rithesh Kumar, Shubham Jain,
  Jose Sotelo, Aaron Courville, and Yoshua Bengio.
\newblock {SampleRNN}: {An} {Unconditional} {End}-to-{End} {Neural} {Audio}
  {Generation} {Model}.
\newblock \emph{arXiv preprint arXiv:1612.07837}, 2016.

\bibitem[Oord et~al.(2018)Oord, Li, Babuschkin, Simonyan, Vinyals, Kavukcuoglu,
  Driessche, Lockhart, Cobo, Stimberg, et~al.]{oord2018parallel}
Aaron Oord, Yazhe Li, Igor Babuschkin, Karen Simonyan, Oriol Vinyals, Koray
  Kavukcuoglu, George Driessche, Edward Lockhart, Luis Cobo, Florian Stimberg,
  et~al.
\newblock Parallel {WaveNet}: {Fast} {High}-{Fidelity} {Speech} {Synthesis}.
\newblock In \emph{International conference on machine learning}, pages
  3918--3926. PMLR, 2018.

\bibitem[Engel et~al.(2017)Engel, Resnick, Roberts, Dieleman, Norouzi, Eck, and
  Simonyan]{engel2017neural}
Jesse Engel, Cinjon Resnick, Adam Roberts, Sander Dieleman, Mohammad Norouzi,
  Douglas Eck, and Karen Simonyan.
\newblock {Neural} {Audio} {Synthesis} of {Musical} {Notes} with {WaveNet}
  {Autoencoders}.
\newblock In \emph{International Conference on Machine Learning}, pages
  1068--1077. PMLR, 2017.

\bibitem[Kumar et~al.(2019)Kumar, Kumar, de~Boissiere, Gestin, Teoh, Sotelo,
  de~Br{\'e}bisson, Bengio, and Courville]{kumar2019melgan}
Kundan Kumar, Rithesh Kumar, Thibault de~Boissiere, Lucas Gestin, Wei~Zhen
  Teoh, Jose Sotelo, Alexandre de~Br{\'e}bisson, Yoshua Bengio, and Aaron~C
  Courville.
\newblock {MelGAN}: {Generative} {Adversarial} {Networks} for {Conditional}
  {Waveform} {Synthesis}.
\newblock \emph{Advances in neural information processing systems}, 32, 2019.

\bibitem[D{\'e}fossez et~al.(2018)D{\'e}fossez, Zeghidour, Usunier, Bottou, and
  Bach]{defossez2018sing}
Alexandre D{\'e}fossez, Neil Zeghidour, Nicolas Usunier, L{\'e}on Bottou, and
  Francis Bach.
\newblock {SING}: {Symbol}-to-{Instrument} {Neural} {Generator}.
\newblock \emph{Advances in neural information processing systems}, 31, 2018.

\bibitem[Caillon and Esling(2021)]{caillon2021rave}
Antoine Caillon and Philippe Esling.
\newblock {RAVE}: {A} variational autoencoder for fast and high-quality neural
  audio synthesis.
\newblock \emph{arXiv preprint arXiv:2111.05011}, 2021.

\bibitem[Zeghidour et~al.(2021)Zeghidour, Luebs, Omran, Skoglund, and
  Tagliasacchi]{zeghidour2021soundstream}
Neil Zeghidour, Alejandro Luebs, Ahmed Omran, Jan Skoglund, and Marco
  Tagliasacchi.
\newblock {SoundStream}: {An} {End}-to-{End} {Neural} {Audio} {Codec}.
\newblock \emph{IEEE/ACM Transactions on Audio, Speech, and Language
  Processing}, 30:\penalty0 495--507, 2021.

\bibitem[Clevert et~al.(2015)Clevert, Unterthiner, and
  Hochreiter]{clevert2015fast}
Djork-Arn{\'e} Clevert, Thomas Unterthiner, and Sepp Hochreiter.
\newblock Fast and {Accurate} {Deep} {Network} {Learning} by {Exponential}
  {Linear} {Units} ({ELUs}).
\newblock \emph{arXiv preprint arXiv:1511.07289}, 2015.

\bibitem[Girshick(2015)]{girshick2015fast}
Ross Girshick.
\newblock Fast {R}-{CNN}.
\newblock In \emph{Proceedings of the IEEE international conference on computer
  vision}, pages 1440--1448, 2015.

\bibitem[Yamamoto et~al.(2020)Yamamoto, Song, and Kim]{yamamoto2020parallel}
Ryuichi Yamamoto, Eunwoo Song, and Jae-Min Kim.
\newblock {Parallel} {WaveGAN}: {A} fast waveform generation model based on
  generative adversarial networks with multi-resolution spectrogram.
\newblock In \emph{ICASSP 2020-2020 IEEE International Conference on Acoustics,
  Speech and Signal Processing (ICASSP)}, pages 6199--6203. IEEE, 2020.

\bibitem[Loshchilov and Hutter(2017)]{loshchilov2017decoupled}
Ilya Loshchilov and Frank Hutter.
\newblock {Decoupled} {Weight} {Decay} {Regularization}.
\newblock \emph{arXiv preprint arXiv:1711.05101}, 2017.

\bibitem[Liu et~al.(2022)Liu, Choi, Liu, Kong, Tian, and Wang]{liu2022neural}
Haohe Liu, Woosung Choi, Xubo Liu, Qiuqiang Kong, Qiao Tian, and DeLiang Wang.
\newblock {Neural} {Vocoder} is {All} {You} {Need} for {Speech}
  {Super}-resolution.
\newblock \emph{arXiv preprint arXiv:2203.14941}, 2022.

\bibitem[Luo and Mesgarani(2018)]{luo2018tasnet}
Yi~Luo and Nima Mesgarani.
\newblock {TasNet}: time-domain audio separation network for real-time,
  single-channel speech separation.
\newblock In \emph{2018 IEEE International Conference on Acoustics, Speech and
  Signal Processing (ICASSP)}, pages 696--700. IEEE, 2018.

\bibitem[Rix et~al.(2001)Rix, Beerends, Hollier, and
  Hekstra]{rix2001perceptual}
Antony~W Rix, John~G Beerends, Michael~P Hollier, and Andries~P Hekstra.
\newblock Perceptual evaluation of speech quality ({PESQ})-a new method for
  speech quality assessment of telephone networks and codecs.
\newblock In \emph{2001 IEEE international conference on acoustics, speech, and
  signal processing. Proceedings (Cat. No. 01CH37221)}, volume~2, pages
  749--752. IEEE, 2001.

\bibitem[Taal et~al.(2010)Taal, Hendriks, Heusdens, and Jensen]{taal2010short}
Cees~H Taal, Richard~C Hendriks, Richard Heusdens, and Jesper Jensen.
\newblock A short-time objective intelligibility measure for time-frequency
  weighted noisy speech.
\newblock In \emph{2010 IEEE international conference on acoustics, speech and
  signal processing}, pages 4214--4217. IEEE, 2010.

\bibitem[Manocha et~al.(2021)Manocha, Jin, Zhang, and
  Finkelstein]{manocha2021cdpam}
Pranay Manocha, Zeyu Jin, Richard Zhang, and Adam Finkelstein.
\newblock {CDPAM}: {Contrastive} learning for perceptual audio similarity.
\newblock In \emph{ICASSP 2021-2021 IEEE International Conference on Acoustics,
  Speech and Signal Processing (ICASSP)}, pages 196--200. IEEE, 2021.

\end{thebibliography}
}

\end{document}